\documentclass[
 aps,
 prd,
 reprint,
 superscriptaddress,
 nofootinbib,
 amsmath,amssymb,
 floatfix,
]{revtex4-2}

\usepackage[T1]{fontenc}
\usepackage[utf8]{inputenc}

\usepackage{newtxtext}
\usepackage{newtxmath}

\DeclareUnicodeCharacter{00C9}{\'E}
\DeclareUnicodeCharacter{00E3}{\~a}
\DeclareUnicodeCharacter{00E9}{\'e}
\DeclareUnicodeCharacter{00F1}{\~n}
\DeclareUnicodeCharacter{00F3}{\'o}
\DeclareUnicodeCharacter{0394}{\ensuremath{\Delta}}
\DeclareUnicodeCharacter{2019}{'}

\DeclareRobustCommand{\VAN}[3]{#2}
\let\VANthebibliography\thebibliography
\def\thebibliography{\DeclareRobustCommand{\VAN}[3]{##3}\VANthebibliography}

\usepackage{graphicx}	
\usepackage{amsmath}	
\usepackage{comment}
\usepackage{bm}
\usepackage{float}
\usepackage{xcolor}

\newcommand{\fnl}{f_{\rm NL}}
\newcommand{\fnld}{f_{\rm NL, \Delta}}\newcommand{\fnldfid}{f_{\rm NL, \Delta}^{\rm fid}}

\usepackage[
    hidelinks=true,
    colorlinks=true,
    linkcolor=blue,
    citecolor=blue,
    urlcolor=blue
]{hyperref}

\AtBeginDocument{%
}

\providecommand{\bsp}{}

\let\k\relax\newcommand{\k}{\mathbf{k}}

\begin{document}
\label{firstpage}


\title{Searching for signatures of inflationary massive fields in DESI Imaging data\\ and Stage-V galaxy surveys}

\author{Walter Riquelme}
\email[Corresponding author: {}]{walter.riquelme@ictp-saifr.org}
\affiliation{ICTP South American Institute for Fundamental Research, IFT-UNESP, S\~{a}o Paulo, SP 01440-070, Brazil}

\author{Santiago Avila}
\affiliation{Centro de Investigaciones Energ\'eticas, Medioambientales y Tecnol\'ogicas (CIEMAT), Madrid, Spain}

\author{Lucas Pinol}
\affiliation{Laboratoire de Physique de l’École Normale Supérieure, ENS, CNRS, Université PSL, Sorbonne Université, Université Paris Cité, F-75005, Paris, France}

\author{Hui Kong}
\affiliation{Institut de F\'{i}sica d’Altes Energies (IFAE), The Barcelona Institute of Science and Technology,\\ Campus UAB, 08193 Bellaterra Barcelona, Spain}

\author{Anna Porredon}
\affiliation{Centro de Investigaciones Energ\'eticas, Medioambientales y Tecnol\'ogicas (CIEMAT), Madrid, Spain}

\date{\today}

\begin{abstract}
We investigate the cosmological imprints of massive fields during inflation through primordial non-Gaussianity (PNG). When these fields are sufficiently light, they produce a signal in galaxy clustering, $\propto f_{\rm NL,\Delta}k^{\Delta-2}$ with $\Delta\in(0,3/2]$, corresponding to a beyond-local PNG contribution to the scale-dependent bias. We use the angular correlation function to constrain $f_{\rm NL,\Delta}$ and $\Delta$ using imaging data used for the targeting of the Dark Energy Spectroscopic Instrument (DESI). For the most aggressive systematics treatment, there is no evidence for local PNG, hence no constraint on $\Delta$. However, when considering a less aggressive treatment, a hint is found with $f_{\rm NL}^{\rm loc}=27^{+10}_{-9}$, consistent with previous analyses. For beyond-local PNG, that signal gives a preference for $f_{\rm NL, \Delta}=5.12^{+6.13}_{-3.62}\times10^{3}$ and $\Delta=0.91^{+0.25}_{-0.19}$. This preference is not robust under decontamination choices, likely driven by residual systematics, and we present it as a showcase for future constraints. Additionally, we forecast the sensitivity of upcoming Stage-V surveys, the Wide-field Spectroscopic Telescope (WST), the MUltiplexed Survey Telescope (MUST), and the Spectroscopic Stage 5 Experiment (Spec-S5), to constrain $\Delta$ using Lyman-break galaxies. Around the local limit, $\Delta^{\rm fid}=0$, we find that they can reach uncertainties of $\sigma(\Delta)\simeq0.17-0.50$, depending on the survey, for a fiducial $f_{\rm NL,\Delta}^{\rm fid}=4$. The constraining power on $\Delta$ increases as we increase $f_{\rm NL,\Delta}^{\rm fid}$ and decreases for larger fiducial $\Delta$. Finally, we derive a relation between the detectability of $\Delta$ and the local PNG constraints, $\sigma(f_{\rm NL}^{\rm loc})$. This provides a tool to estimate the survey sensitivity required to resolve massive-field signatures.
\end{abstract}
\maketitle

\section{Introduction}

Cosmic inflation predicts that the primordial seeds of structure are described by close to Gaussian random fields, for which most of the statistical information is contained in the two-point function.
Departures from Gaussianity, also known as primordial non-Gaussianity (PNG), provide a way to distinguish between different inflationary scenarios.
The leading statistic used to describe PNG is the primordial bispectrum, or three-point correlation function.
A commonly used template is the local PNG, defined through a local quadratic correction to the primordial potential,
\begin{equation}
    \Phi(\mathbf{x})
    =
    \phi_G(\mathbf{x})
    +
    \fnl^{\rm loc}
    \left[
        \phi_G^2(\mathbf{x})
        -
        \langle \phi_G^2\rangle
    \right],
\end{equation}
where $\phi_G$ is a Gaussian field and $\fnl^{\rm loc}$ sets the amplitude of the non-Gaussian correction \cite{2001PhRvD..63f3002K}.
This template generates a bispectrum that peaks in the squeezed limit, where one mode has a much longer wavelength than the other two.
For canonical single-field attractor inflation, the physically observable squeezed limit vanishes due to gauge artifact cancellation, yielding zero local PNG \cite{2011JCAP...05..014T, 2013PhRvD..88h3502P}.
Therefore, a detection of $\fnl^{\rm loc}$ would point to additional degrees of freedom or to non-standard inflationary dynamics.

Local PNG is especially relevant for the Large-Scale Structure (LSS) of the Universe because it induces a scale-dependent correction to the galaxy bias.
This scale-dependent bias produces an enhancement of galaxy clustering on very large scales and can therefore be probed using two-point statistics of biased tracers.
The effect has been extensively studied as a probe of local PNG \cite{2008PhRvD..77l3514D, 2008JCAP...08..031S, 2008ApJ...677L..77M}.

In parallel, the Dark Energy Task Force introduced the classification of dark energy galaxy surveys into different stages \cite{2006astro.ph..9591A}.
Although this classification was originally motivated by dark energy, these surveys have also become key for tests of inflation with large-scale structure.
In particular, Stage-III surveys data sets have been used to constrain local PNG through the scale-dependent bias in the clustering of galaxies \cite{2013MNRAS.428.1116R, 2014PhRvD..89b3511G, 2015JCAP...05..040H, 2014PhRvL.113v1301L, 2019JCAP...09..010C, 2022MNRAS.514.3396M}.

Current Stage-IV data are already improving the constraints on local PNG. 
The Dark Energy Spectroscopic Instrument (DESI) \cite{2016arXiv161100036D} is a Stage-IV spectroscopic survey designed to provide precise measurements of LSS and the properties of our Universe. 
The DESI Legacy Imaging data \cite{2019AJ....157..168D}, hereafter DESI Imaging, are a collection of photometric galaxy samples from different surveys, used to define the targets for DESI galaxy and quasar spectroscopic samples.
Recently, DESI Imaging showed a preference for non-zero $\fnl^{\rm loc}$ in the angular clustering of Luminous Red Galaxies (LRGs), although this signal is likely affected by residual observational systematics \cite{2024MNRAS.532.1902R}.
At the same time, recent DESI spectroscopic constraints are consistent with $\fnl^{\rm loc}=0$, highlighting the importance of systematics control in PNG measurements from galaxy clustering \cite{2025JCAP...06..029C, 2026arXiv260405213R, 2026JCAP...05..098C, 2026PhRvD.113f3552C}.

Future galaxy surveys are expected to significantly improve the sensitivity to local PNG.
Proposed Stage-V surveys such as the Wide-field Spectroscopic Telescope (WST) \cite{2024arXiv240305398M}, the MUltiplexed Survey Telescope (MUST) \cite{2024arXiv241107970Z}, and the Spectroscopic Stage 5 Experiment (Spec-S5) \cite{2025arXiv250307923B} will target high-redshift Lyman-Break Galaxies (LBGs) \cite{2024JCAP...08..059R,2025MNRAS.543.3196Y,2025ApJS..281...54C}, combining large survey volumes, high galaxy bias, and large sky coverage.
These properties make LBGs a potentially powerful sample for PNG measurements \cite{2026A&A...709A..70P}.
Forecasts for future spectroscopic surveys suggest that they could surpass current constraints from temperature fluctuations of the Cosmic Microwave Background (CMB) \cite{2020A&A...641A...9P} and reach $\sigma(\fnl^{\rm loc})\lesssim1$ \cite{2023MNRAS.521.3648D}.
If such surveys detect a non-zero local PNG amplitude, an important question is whether the signal can be uniquely associated with the local PNG template, or whether it could arise from a more general squeezed limit signal. This motivates going beyond the local PNG template.

Additional fields present during inflation can also generate correlations between long- and short-wavelength modes in the squeezed limit, but with a scale dependence that differs from the local ansatz \cite{2010JCAP...04..027C, 2013JHEP...10..171G}.
These signals can still induce scale-dependent bias in galaxy clustering, with a signal controlled by both the PNG amplitude and the scaling of the squeezed limit.
Early studies showed that the galaxy power spectrum can be used to probe quasi-single-field and more general squeezed-limit PNG through the scale-dependent bias \cite{2012JCAP...08..019N, 2012JCAP...08..033S}.
This idea was later developed for broader classes of PNG beyond the local type, including the role of galaxy bias degeneracies and the incorporation of multi-tracer techniques \cite{2017JCAP...04..002G}.
A related phenomenological search was applied to the eighth data release of the Sloan Digital Sky Survey (SDSS-DR8) in Ref.~\cite{2014JCAP...02..038A}.
In this work, we follow the massive-field parameterization of Ref.~\cite{2024JCAP...05..090G}, where the departure from the local limit is described by the scaling parameter $\Delta$ related to the mass of the primordial fields. 

In this paper, we study how these signatures appear in current DESI Imaging data and in forecasts for future Stage-V surveys.
We first apply the angular correlation function (ACF) with scale-dependent bias pipeline developed in Ref.~\cite{2023MNRAS.523..603R}, hereafter Riquelme23, to the DESI Imaging LRG sample \cite{2023AJ....165...58Z}.
This allows us to test whether the ACF pipeline reproduces the local PNG constraints found in the harmonic space analysis of Ref.~\cite{2024MNRAS.532.1902R}, hereafter Rezaie24.
We then extend the theoretical model to include the parameter $\Delta$ and study how an apparent local PNG signal projects onto the signatures of massive fields.
Here, the DESI Imaging measurement is used as a real-data test case for the extended model, rather than as evidence for a beyond-local PNG signal.
We also forecast the constraining power of massive-field signatures on Stage-V spectroscopic surveys using high-redshift LBGs samples.
For WST, MUST, and Spec-S5, we first quantify the baseline sensitivity to local PNG and then study whether this sensitivity can be translated into constraints on $\Delta$.
This allows us to determine when future surveys can distinguish the local limit from a more general beyond-local scale dependence.
Together, these analyses connect the detectability of local PNG with the ability to resolve departures from the local limit.

The paper is organized as follows.
In \autoref{sec:theory}, we describe the theoretical framework of massive fields during inflation, their imprint on galaxy clustering, and our statistical methodology.
\autoref{sec:Data} describes the DESI Legacy Imaging data and the specifications of the upcoming Stage-V surveys.
In \autoref{sec:delta_desi}, we show the constraints derived from the DESI data, while in \autoref{sec:delta_stagev}, we provide forecasts for future surveys.
Last, we present our concluding remarks in \autoref{sec:Conclusions}.

\section{Theory and methods}\label{sec:theory}

In this section, we describe the massive-field model, its imprint on the power spectrum and ACF, and the parameter-inference framework.

\subsection{Massive fields during inflation}

We define the bispectrum of the primordial curvature perturbation $\zeta$ as
\begin{equation}
    \langle \zeta(\k_{1})\zeta(\k_{2})\zeta(\k_{3}) \rangle=  (2\pi)^3 \delta^{(3)}(\k_{1}+\k_{2}+\k_{3})B_\zeta(k_1,k_2,k_3)\,,
\end{equation}
where statistical isotropy enforces wavevectors to form a closed triangle in Fourier space and statistical homogeneity explains the dependence of the bispectrum on the moduli $k_i=|\k_{i}|$ only.
The generic template of the bispectrum for the mixing between a massive field $\sigma$ of mass $m$ and the curvature perturbations $\zeta$ during inflation takes a universal form in the squeezed limit, with $\k_{1}=\k_{S}+\k_{L}/2\,,\,\k_{2}=-\k_{S}+\k_{L}/2\,,\,\k_{3}=-\k_{L}$ with $k_L \ll k_S$,
\begin{equation}
    B_\zeta(k_1,k_2,k_3) \overset{\text{squeezed}}{=} \frac{12}{5} f_{\rm NL, \Delta} \left(\frac{k_S}{k_L}\right)^\Delta P_\zeta(k_S) P_\zeta(k_L) + \ldots \,,
\end{equation}
where the dots indicate terms at higher order in $k_L/k_S$.
The quantity $f_{\rm NL,\Delta}$ corresponds to the amplitude of primordial non-Gaussianity, which depends on the regime of parameters and interactions during inflation.
The scaling of the power-law squeezed-limit signal for massive fields in the mass range of $m<3H/2$ is given by \cite{2010JCAP...04..027C}
\begin{equation}\label{eq:delta}
    \Delta = \frac{3}{2}
    -
    \sqrt{
    \frac{9}{4}
    -
    \frac{m^2}{H^2}
    },
\end{equation}
where $H$ is the Hubble scale during inflation.

The local PNG limit is recovered for $\Delta=0$ (corresponding to $m=0$), for which we recover the usual amplitude $f_{\rm NL}^{\rm loc} =f_{\rm NL,\Delta =0}$.
Very small values of $\Delta$ are therefore close to the local shape, while larger values explore different masses of the extra fields.
For example, $\Delta=0.4$ and $\Delta=1.0$ correspond to fields with masses $m\simeq H$ and $m=\sqrt{2}H$, respectively.
The case $m\sim H$ is characteristic of the quasi-single-field regime, where massive isocurvature fields can generate intermediate squeezed-limit shapes \cite{2010JCAP...04..027C}.
The value $m=\sqrt{2}H$ corresponds to the conformally coupled scalar field in de Sitter space \cite{Arkani-Hamed:2018kmz}.

The primordial bispectrum affects late-time galaxy clustering through its squeezed configuration.
In this limit, the long wavelength mode $k_L$, later associated with the large scale galaxy clustering, is coupled to the short wavelength modes $k_S$, which control the local amplitude of small scale fluctuations through the amplitude of the squeezed primordial bispectrum, $\fnld$.
This coupling allows the large-scale mode to change the local abundance of tracers, producing a scale-dependent contribution to the galaxy bias \cite{2018PhR...733....1D}. 

The short modes are associated with a smoothing scale $R\sim 1/k_S$.
Following Ref.~\cite{2024JCAP...05..090G}, we take $R=R^\ast\simeq2.66\,h^{-1}{\rm Mpc}$, corresponding to a minimum halo mass of $10^{13}M_\odot$.
This choice provides a common reference scale for the short modes entering the squeezed limit.
Since the large-scale modes used in our analysis satisfy $kR^\ast<1$, the long modes remain separated from the short modes associated with halo formation.
For lower-mass tracers, the corresponding Lagrangian radius would be smaller, making this scale separation even safer.

For massive fields during inflation, as shown before, the squeezed bispectrum has an additional scale dependence controlled by $\Delta$.
This scaling is transferred to the scale-dependent bias as \cite{2009ApJ...706L..91V, 2013JCAP...05..001B, 2015JCAP...12..043A}
\begin{equation} \label{eq:scale_dep}
    b(k,z) = b + \fnld (kR^{*})^{\Delta} 2\delta_{c}(b-p) \alpha(k, z),
\end{equation}
where $b$ is the galaxy linear bias, $\delta_{c}=1.686$ is the critical value of collapse for halo formation in an Einstein-de Sitter universe \cite{1984ApJ...281....1F}, and $\alpha(k,z)$ is given by
\begin{equation}\label{eq:alpha}
    \alpha(k,z)=\frac{3\Omega_{\rm m}}{2D(z)}\frac{H_{0}^{2}}{c^{2}}\frac{g(0)}{g(z_{\rm rad})}\frac{1}{k^{2}T(k)}.
\end{equation}
Here, $H_{0}$ is the Hubble constant today, $c$ is the speed of light, and $\Omega_{\rm m}$ is the matter density parameter today.
In addition, $T(k)$ is the transfer function and $D(z)$ is the linear growth factor, normalized to unity at $k=0$ and $z=0$, respectively.
The factor $g(0)/g(z_{\rm rad})\approx1.3$, with $g(z)=(1+z)D(z)$, appears because $D(z)$ is normalized to unity today, and can be omitted if the growth factor is instead normalized to the scale factor during matter domination \cite{2019MNRAS.485.4160M}.
For our analysis, we assume the universality relation for the halo mass function and fix $p=1$ \cite{2008JCAP...08..031S, 2008PhRvD..77l3514D}.

\subsection{Galaxy Power Spectrum}
The scale-dependent galaxy bias modifies the observed redshift-space galaxy power spectrum (PS), which can be modeled as 
\begin{equation}
    P_{\rm g}(k, \mu, z) = \left( b(k, z) + f(z)\mu^2 \right)^2 P_{\rm m}(k, z),
\end{equation}
where $P_{\rm m}(k, z)$ is the underlying matter power spectrum. 
The matter power spectrum and transfer function are computed with
\texttt{CAMB} \cite{2000ApJ...538..473L, 2012JCAP...04..027H}.
We also include Redshift-Space Distortions (RSD) on linear scales using the Kaiser effect where $f(z)\approx\Omega_{\rm m}(z)^{\gamma}$ is the growth rate of structures with $\gamma=0.55$ \cite{2005PhRvD..72d3529L}, and $\mu$ is the cosine of the angle between the wavevector $\k$ and the line of sight $\hat{\textbf{l}}$.

\begin{figure}
    \includegraphics[width=\columnwidth]{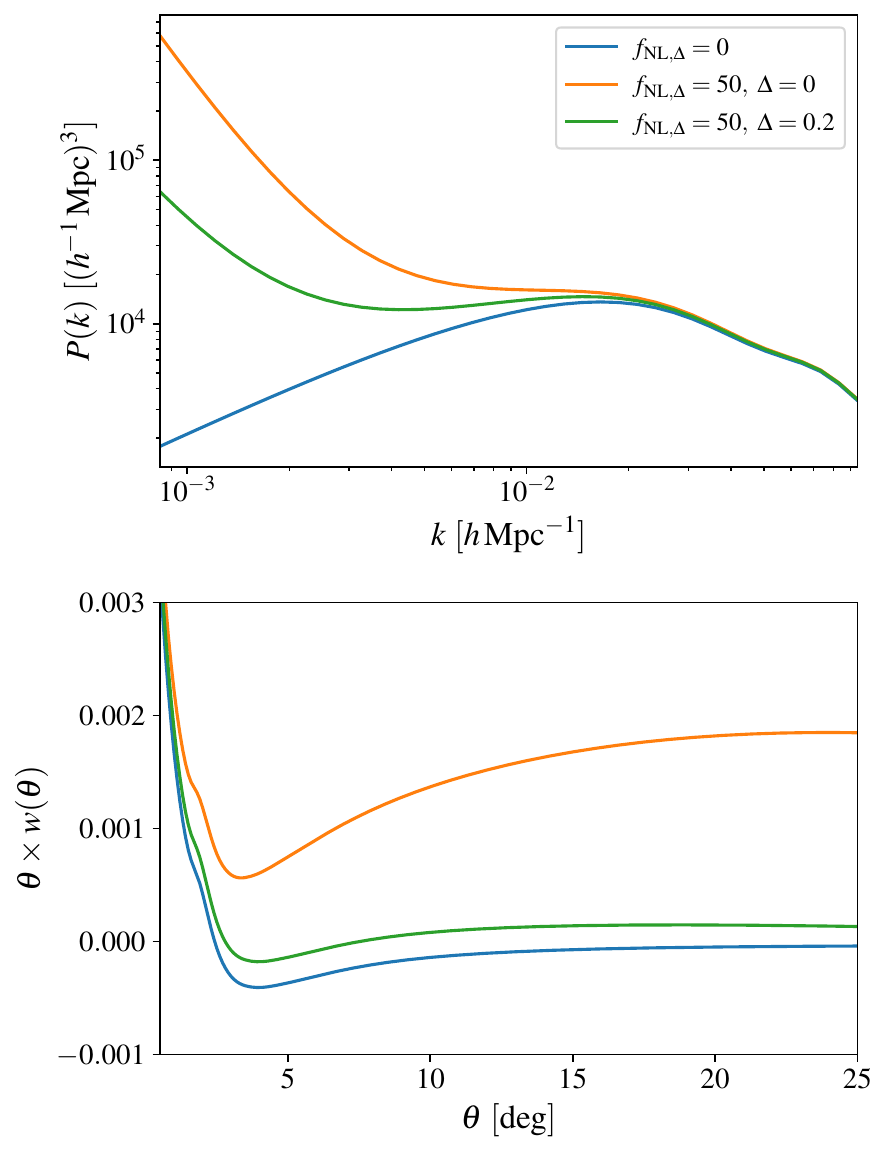}
    \caption{\textbf{Top panel:} Theoretical galaxy power spectrum for different values of both $\fnld$ and $\Delta$. \textbf{Bottom panel:} Angular correlation function for the DESI Imaging as function of $\fnld$ and $\Delta$. The ACF also includes the effect of the integral constraint. The figure shows how $\fnld$ increases the large-scale clustering signal and how it is suppressed as $\Delta$ increases.}
    \label{fig:wtheta_theory}
\end{figure}

In the top panel of \autoref{fig:wtheta_theory}, we show the galaxy power spectrum for different values of both $\fnld$ and $\Delta$. The figure shows the large-scale enhancement effect induced by the scale-dependent bias as $\fnld$ increases. Also, as we increase the value of $\Delta$, we can see that the enhancement is suppressed. This degeneracy is expected given the relation between $\fnld$ and $\Delta$ from \autoref{eq:scale_dep}, and corresponds to the main source of difficulty in putting constraints on $\Delta$.

\subsection{Angular correlation function}

For the photometric DESI Imaging sample, we use the projected angular correlation function.
Using the previously described galaxy power spectrum, we can compute its configuration space counterpart, the two-point correlation function (2PCF),
\begin{equation}\label{eq:2PCF}
\xi(r, \hat{\textbf{r}}\cdot\hat{\textbf{l}}) = \sum_{\ell=0,2,4}\frac{i^{\ell}}{2\pi^{2}} 
\int_{0}^{\infty}\text{d}k\ k^{2}j_{\ell}(k r)P_{\ell}(k, z)L_{\ell}(\hat{\textbf{r}}\cdot\hat{\textbf{l}})
\end{equation}
with
\begin{equation}
P_{\ell}(k, z) \equiv \frac{(2\ell+1)}{2}\int_{-1}^{1}\text{d}\mu\ L_{\ell}(\mu)P_{\rm g}(k, \mu, z),  
\end{equation}
also, $r$ is the separation distance between galaxies, $j_{\ell}$ is the spherical Bessel function, and $L_{\ell}$ is the Legendre polynomial.
The correlation function is also a function of the angle between the line of sight direction $\hat{\textbf{l}}$ and the direction of the separation vector $\hat{\textbf{r}}$, given by
\begin{equation}
    \hat{\textbf{r}}\cdot\hat{\textbf{l}} = \frac{\chi(z_{2})-\chi(z_{1})}{r}\cos{\frac{\theta}{2}},    
\end{equation}
where $\chi(z)$ is the comoving distance, and $\theta$ is the angular separation between two galaxies.
We compute the ACF as
\begin{equation} \label{eq:acf_fnl}
    w_{\rm raw}(\theta) = \int\mathrm{d}z_{1}\int\mathrm{d}z_{2}\, \phi(z_1)\phi(z_2)\xi(r(z_1, z_2, \theta), \hat{\textbf{r}}\cdot\hat{\textbf{l}}),
\end{equation}
where $\phi(z)=N(z)D(z)$, with $N(z)$ the galaxy redshift distribution normalized to unity.
The separation between two galaxies as a function of the angular separation is defined through the relation
\begin{equation}
    r(z_{1}, z_{2}, \theta) = \left(\chi(z_{1})^{2} + \chi(z_{2})^{2} - 2\chi(z_{1})\chi(z_{2})\cos{\theta}\right)^{1/2}.
\end{equation}

One property of the measured ACF is that, since the mean galaxy density is estimated from the finite survey footprint, its integral over the footprint vanishes. This is known as the integral constraint.
Since PNG induces an enhancement of clustering at large angular scales, this correction must be applied consistently to the theoretical prediction to avoid biased constraints, as shown in Riquelme23.
We compute it from the uncorrected theory ACF,
\begin{equation}
    I(\fnld,\Delta)
    =
    \frac{\sum^{\theta_{\rm lim}} RR(\theta)\,
    w_{\rm raw}(\theta,\fnld,\Delta)}
    {\sum^{\theta_{\rm lim}} RR(\theta)},
    \label{eq:ic_thetamax}
\end{equation}
where $RR(\theta)$ is the number of random-random pairs at angular separation $\theta$ for the DESI Imaging angular mask. 
A relevant point in the computation of the integral constraint is that $\theta_{\rm lim}$ is not the maximum angular scale used in the measurement, but the maximum angular separation allowed by the survey mask.
Therefore, even though our measurement is performed up to $\theta_{\rm max}=20\deg$, the integral constraint is computed using $w_{\rm raw}(\theta)$ up to $\theta_{\rm lim}\sim 140\deg$.
We can then model the full theoretical ACF as
\begin{equation}\label{eq:acf_full}
    w(\theta,\fnld,\Delta) = w_{\rm raw}(\theta,\fnld,\Delta) - I(\fnld,\Delta).
\end{equation}
The bottom panel of \autoref{fig:wtheta_theory} shows the theoretical ACF computed from \autoref{eq:acf_full} for different values of $\fnld$ and $\Delta$. Similar to the galaxy power spectrum, $\fnld$ enhances the ACF at large angular separations, while increasing $\Delta$ suppresses this effect.

\subsection{Parameter inference}\label{sec:inference}

In this section, we describe the statistical methods used to constrain signatures of primordial massive fields using galaxy clustering.
We constrain $\Theta=\{\fnld,\Delta,b\}$ using Gaussian likelihoods for the DESI Imaging ACF and Stage-V power-spectrum forecasts, varying the parameters jointly to capture their possible degeneracies.
As typically done in PNG analyses, throughout this work, we assume flat-$\Lambda$CDM with fixed cosmological parameters\footnote{For the DESI Imaging pipeline, we adopt the density parameter for matter $\Omega_{\rm m}=0.25$, for baryons $\Omega_{\rm b}=0.044$, the scalar spectral index $n_{\rm s}=0.95$, the dimensionless Hubble constant $h=0.7$, the amplitude of the power spectrum at $8h^{-1}\text{Mpc}$, $\sigma_8=0.8$, and the sum of neutrino masses $M_\nu=0$. For the Stage-V forecasts, we adopt values consistent with the Planck 2018 baseline \cite{2020A&A...641A...6P}: $\Omega_{\rm m} = 0.31$, $\Omega_{\rm b} = 0.048$, $n_{\rm s} = 0.9675$, $h = 0.68$, the scalar amplitude of primordial fluctuations $A_{\rm s} = 2.1 \times 10^{-9}$, and $M_\nu=0$}. 

For a given observable, we define a data vector $\mathbf{d}$ and a theoretical model vector $\mathbf{m}(\Theta)$. The likelihood is then written as
\begin{equation}
    \ln \mathcal{L}(\mathbf{d}|\Theta)
    =
    -\frac{1}{2}\chi^{2}(\Theta),
    \label{eq:general_likelihood}
\end{equation}
with
\begin{equation}
\chi^{2}(\Theta)=
    \left[
    \mathbf{m}(\Theta)-\mathbf{d}
    \right]^T
    \mathbf{C}^{-1}
    \left[
    \mathbf{m}(\Theta)-\mathbf{d}
    \right],
    \label{eq:chi2}
\end{equation}
where $\mathbf{C}$ is the covariance matrix associated with the observable.
For both ACF and PS analyses, the covariance is evaluated using a reference model with $f_{\rm NL}=0$ and the fiducial bias values shown in the fifth column of \autoref{tab:surveys}, and is kept fixed during the parameter inference.

Using Bayes' theorem, the posterior distribution is
\begin{equation}
    P(\Theta|\mathbf{d})=\frac{\mathcal{L}(\mathbf{d}|\Theta)\,\Pi(\Theta)}{\mathcal{Z}},
    \label{eq:posterior}
\end{equation}
where $\Pi(\Theta)$ is the prior distribution and $\mathcal{Z}$ is the Bayesian evidence.
The marginalized posterior for a parameter $\Theta_i$ is then obtained by integrating over the remaining parameters,
\begin{equation}
    P(\Theta_i|\mathbf{d})=
    \int \mathrm{d}\Theta_{\neq i}\,P(\Theta|\mathbf{d}).
\end{equation}

For the DESI Imaging sample, the data vector is the observed angular correlation function, $\mathbf{d}_{\rm ACF}=\left\{w^{\rm obs}(\theta)\right\}$, as described in \autoref{sec:desi_data}.
The likelihood for the ACF is then given by \autoref{eq:general_likelihood} with $\mathbf{m}(\Theta)=w(\theta|\Theta)$.
The covariance matrix of the angular correlation function is computed analytically with \texttt{CosmoCov} \cite{2017MNRAS.470.2100K, 2020MNRAS.497.2699F}, assuming a Gaussian covariance that includes sample variance and Poisson shot noise.

For the Stage-V power spectrum forecast, the data vector is constructed from the theoretical galaxy power spectrum evaluated at fixed fiducial values, $\Theta^{\rm fid}=\{\fnld^{{\rm fid}}, \Delta^{\rm fid},b^{\rm fid}\}$, as $\mathbf{d}_{\rm PS}=\left\{P_{\rm g}(k,\mu,z|\Theta^{\rm fid})\right\}.$
On the scales used in this analysis, the Fourier modes are treated as statistically independent. The covariance is therefore diagonal, and \autoref{eq:chi2} reduces to
\begin{equation}
\chi^{2}_{\rm PS}(\Theta)=
    \sum_i\sum_k\sum_\mu
    \frac{\left(P_{\rm g}(k,\mu,z_i|\Theta)-P_{\rm g}(k,\mu,z_i|\Theta^{\rm fid})\right)^2}
    {\sigma^2_{\rm PS}(k,\mu,z_i)}.
    \label{eq:chi2_PS}
\end{equation}
Since the data vector is generated from the same model used in the fit, the likelihood is maximized at the fiducial parameter values by construction. The goal is to quantify the expected constraining power and parameter degeneracies around these values.
As in the ACF case, the variance of the power spectrum measurement is determined by the combination of cosmic variance and Poisson shot noise.
For a discrete grid in $k$ and $\mu$, the variance is given by \cite{2003ApJ...598..720S}
\begin{equation}
    \sigma^2_{\rm PS}(k,\mu,z_i)
    =
    \frac{8 \pi^2}{V_{\rm bin} k^2 \Delta k \Delta \mu}
    \left(
    P_{\rm g}(k,\mu,z_i) + \frac{1}{\bar{n}_{\rm g}}
    \right)^2,
\end{equation}
where $V_{\rm bin}$ is the comoving volume of the survey redshift bin, $\bar{n}_{\rm g}$ is the mean comoving number density of galaxies in each bin, and $\Delta k$ and $\Delta \mu$ are the widths of the Fourier grid. In the variance, $P_{\rm g}$ is evaluated at $f_{\rm NL}=0$ and at the fiducial bias values.
The small-scale cutoff is chosen to keep the analysis within the linear regime where we set $k_{\rm max}=0.1\,h\,{\rm Mpc}^{-1}$.
The large-scale cutoff is set by the transverse survey size at the effective redshift of each bin, $k_{\min,i}=360/(\chi(z_i)\sqrt{A_{\rm survey}})$, with $A_{\rm survey}$ the area of the survey footprint expressed in square degrees. The estimated areas for Stage-V surveys are presented in the fourth column of \autoref{tab:surveys}.
The way to calculate the large-scale cutoff is analogous to the volume-based choices commonly used in PNG forecasts, such as $k_{\min}\sim \pi/V^{1/3}$ \cite{2017PhRvD..95l3513D}.

For both the DESI Imaging measurement and the Stage-V forecast, we consider the following flat prior for the parameter space: $\Delta\in[0, 3/2]$ and $b\in[1.05, 7]$.
The beyond-local PNG constraints are sensitive to the prior range adopted for $\fnld$, especially in cases where the posterior follows the divergent $\fnld$--$\Delta$ direction. Therefore, in order to keep them uninformative, we extend the prior to $\fnld\in[-10^{5}, 10^{7}]$ for the decontaminated DESI Imaging chains, so that the one-dimensional posterior is fully covered. For the forecast, we reduce the prior to $\fnld\in[-10^{3}, 10^{3}]$ considering its expected improvement in the constraining power of the amplitude of PNG.

The posterior distributions are sampled using the importance nested sampling code \texttt{Nautilus} \cite{2023MNRAS.525.3181L} with $n_{\rm live}=80000$.
Using nested sampling methods allows us to better capture non-Gaussian posteriors due to possible degeneracies between $\fnld$ and $\Delta$ \cite{2017JCAP...04..002G}.
We analyze the weighted posterior samples produced by \texttt{Nautilus} using \texttt{GetDist} \cite{2025JCAP...08..025L} to derive the marginalized parameter constraints.
Unless otherwise stated, all the constraints quoted correspond to 1D marginalized $68\%$ credible level (C.L.) around the median of the posterior.
For parameters whose posterior extends to the prior boundary, we quote the corresponding one-sided $95\%$ marginalized credible limit.
We also report posterior probability masses for specific parameter regimes, such as $P(\fnld>0)$, computed directly from the weighted samples.
In the DESI Imaging data analysis, we evaluate the model preference using the $\Delta\chi^2$, with the global minimum identified via a grid-based profile likelihood \cite{2025PhRvD.111h3504H} and \texttt{SciPy}'s differential-evolution optimizer \cite{2020SciPy-NMeth,1997JGOpt..11..341S}, rather than the nested sampler, to ensure a robust estimate of the minimum in highly degenerate likelihoods.

\section{Data}\label{sec:Data}

In this section, we describe the galaxy samples used in our analysis.
We first introduce the DESI Legacy Imaging LRG sample, which is used for the real data measurement.
We then describe the Stage-V samples used for the forecasts.

\begin{table*}
\centering
\setlength{\tabcolsep}{5pt}
\renewcommand{\arraystretch}{1.4}
\caption{Summary of the galaxy sample properties used in this work. The DESI Imaging LRG sample is used for the real data analysis, while the LBG dropout samples define the Stage-V forecast configurations.}
\begin{tabular}{lccccc}
\hline \hline
Survey & Mean redshift & Number density 
$[\mathrm{deg}^{-2}]$ ($\mathrm{arcmin}^{-2}$) & Area $[\mathrm{deg}^{2}]$ & Bias & Reference \\
\hline \hline
DESI Imaging LRG & 0.72 & 792(0.22) & 14,970 & 1.9 & Ref.~\cite{2024MNRAS.532.1902R} \\
\hline
WST, $u$-dropout & 3 & 5,000 (1.39) & 18,000 & 4.5 & Fig. 65 and Sec. 6.6.1 from Ref.~\cite{2024arXiv240305398M} \\
WST, $g$-dropout & 4 & 4,000 (1.11) & 18,000 & 5.3 & \\
WST, $r$-dropout & 5 & 1,500 (0.42) & 18,000 & 6.4 & \\
\hline
MUST, $u$-dropout & 3 & 1,200 (0.33) & 11,000 & 4.5 & Table 3 and Eq. 10 from Ref.~\cite{2024arXiv241107970Z} \\ 
MUST, $g$-dropout & 4 & 800 (0.22) & 11,000 & 5.3 & \\
MUST, $r$-dropout & 5 & 200 (0.06) & 11,000 & 6.4 & \\
\hline
Spec-S5, $u$-dropout & 3 & 1,500 (0.42) & 11,000 & 4.5 & Sec. 4.2.6 from Ref.~\cite{2025arXiv250307923B} \\
Spec-S5, $g$-dropout & 4 & 1,000 (0.28) & 11,000 & 5.3 & \\
\hline
\end{tabular}
\label{tab:surveys}
\end{table*}

\subsection{DESI Legacy Imaging}\label{sec:desi_data}

To constrain the signatures of massive fields in observational data, we use public data from the DESI Legacy Imaging Survey \cite{2019AJ....157..168D}, which provides the target galaxies for DESI spectroscopy.
The data set combines multiple imaging campaigns, including the Mayall $z$-band Legacy Survey \cite{2019AJ....157..168D}, the Beijing Arizona Sky Survey \cite{2017PASP..129f4101Z}, and the Dark Energy Camera Legacy Survey (DECam) \cite{2015AJ....150..150F}, covering a total footprint of $\sim 15,000\,{\rm deg}^{2}$.

For our analysis, we use Luminous Red Galaxies (LRGs).
These galaxies populate massive halos, making them highly biased tracers of the underlying matter density.
They exhibit a strong 4,000\,\AA\ break, which allows for robust photometric redshift estimation.
In particular, we use the public catalog of LRGs selected from DESI Imaging data by Ref.~\cite{2023AJ....165...58Z} and used for the PNG analysis of Rezaie24.
The selected galaxies cover a redshift range of $0.2 < z < 1.35$, with a broad redshift distribution peaked around $z\simeq0.7$ and inferred from early DESI spectroscopy for the imaging LRG sample (see Fig.~1 and Appendix~A3 of Rezaie24).
This redshift distribution is then used as the $N(z)$ when computing the theoretical ACF from \autoref{eq:acf_fnl}.
A summary of the properties of the sample is presented in the first row of \autoref{tab:surveys}.

Because the PNG contribution is strongest on large scales, the analysis is highly sensitive to residual observational systematics.
The systematics in the photometric LRG sample are corrected using weights derived with \texttt{SYSNet} \cite{2020MNRAS.495.1613R,2021MNRAS.506.3439R}, a neural-network-based method that models the dependence of the observed galaxy density on imaging systematics maps at the \texttt{HEALPix} pixel level. 
Following Rezaie24, we consider two sets of nonlinear weights. The first one uses three maps: Galactic extinction, $z$-band depth, and $r$-band PSF size.
We refer to this case as \textit{Nonlinear Three Maps}.
The second one uses nine maps: Galactic extinction, depths in $g$, $r$, $z$, and $W1$, PSF sizes in $g$, $r$, and $z$, and stellar density.
We refer to this case as \textit{Nonlinear Nine Maps}.
The nine-map case removes more residual correlations with imaging properties, but it can also be more affected by overcorrection, since adding more maps can remove part of the true large-scale clustering signal.

To obtain the measured ACF, $w^{\rm obs}(\theta)$, we start from the pixelized map of the contaminated galaxy sample, hereafter labeled as \textit{No weights}, with $N_{\rm side}=256$.
This pixel resolution is sufficient for our analysis since we are interested in large scales.
We then apply the systematic weight maps described above. All of these data were accessed from the public repository associated with the publication of Ref.~\cite{2024MNRAS.532.1902R}.
After applying the systematic weights, we compute the galaxy number density in each pixel and construct the corresponding density contrast map.
The angular correlation function is then estimated from the pixel density contrast using \texttt{TreeCorr} \cite{2004MNRAS.352..338J} with its default two-point correlation settings.
For the measurement, we adopt a linear angular binning with $\Delta \theta=0.4\deg$ and consider three angular ranges: $1\deg<\theta<20\deg$, corresponding to the optimized configuration of Riquelme23; $1\deg<\theta<25\deg$; and $0.5\deg<\theta<25\deg$. The last case extends the measurement down to $\theta_{\rm min}=0.5\deg$, which approximately marks the boundary of the linear regime.

\subsection{Stage-V surveys}\label{sec:stagev_data}

To study how the signatures of massive fields during inflation could appear in future surveys, we first need to define the properties of the expected samples used for the forecast of beyond-local PNG.
Stage-V surveys encompass a collection of spectroscopic surveys that, for our interest, are expected to reach unprecedented constraining power on local PNG, with $\sigma(\fnl^{\rm loc})\lesssim1$.
This makes them an excellent target for searching beyond-local PNG.

One thing that all Stage-V surveys have in common is their potential to capture the spectral properties of galaxies at high redshift, $z>2$.
Since higher redshifts probe larger cosmological volumes, these tracers are expected to be one of the main ways of capturing large-scale galaxy clustering signals, such as local PNG.
In particular, these surveys are being designed to target Lyman-break galaxies (LBGs).
LBGs are young star-forming galaxies characterized by a break in their spectrum at wavelengths shorter than 912\,\AA.
This break can be identified through the absence of flux in different photometric filters.
For example, if the break falls in the $u$-band, around the observed wavelength expected for galaxies at $z\sim3$, these galaxies are labeled as $u$-dropout LBGs.

The Wide-field Spectroscopic Telescope (WST) \cite{2024arXiv240305398M} is a proposed facility with a 12-meter telescope and a $3.1\,{\rm deg}^{2}$ field of view.
It combines multi-object spectroscopy, with a fiber density about 10 times larger than DESI, and panoramic integral field spectroscopy over an expected area of $18,000\,{\rm deg}^{2}$.
As suggested before, one of its main goals is to target LBGs as a tool to measure PNG.
The expected targets are $\{u,g,r\}$-dropouts with unprecedented number density, see for example Fig.~65 of Ref.~\cite{2024arXiv240305398M}.
The summary of the WST properties used in the forecast is shown in the second section of \autoref{tab:surveys}.

The MUltiplexed Survey Telescope (MUST) \cite{2024arXiv241107970Z} is a 6.5-meter telescope under development that will be located in Qinghai, China.
MUST is designed to conduct a spectroscopic survey targeting high-redshift samples across $\sim 13,000\,{\rm deg}^{2}$ in the northern sky.
In particular, for LBGs, it is expected to capture $\{u,g,r\}$-dropouts in an area of $\sim 11,000\,{\rm deg}^{2}$, with estimated number densities of $1200$, $800$, and $200\,{\rm deg}^{-2}$ for $u$-, $g$-, and $r$-dropouts, respectively.
The summary of the MUST properties used in the forecast is shown in the third section of \autoref{tab:surveys}.

The Spectroscopic Stage 5 Experiment (Spec-S5) \cite{2025arXiv250307923B} is a proposed project to be built on the telescope locations of DESI and DECam, using two upgraded 6-meter telescopes and a $3.8\,{\rm deg}^{2}$ area field of view.
The spectrographs are expected to be based on the DESI design, but with upgraded detectors, and each telescope is planned to have about $13,000$ fibers.
One of its main characteristics is that the survey is designed to operate in two different hemispheres, allowing near-full-sky coverage.
Although the survey aims for large sky coverage, its detection of LBGs will be based on targets selected from the Vera C. Rubin's Legacy Survey of Space and Time (LSST) \cite{2019ApJ...873..111I, 2025ApJS..281...54C}, which is located in Chile and will observe the Southern sky.
Since the LBG selection is expected to focus on galaxies within $2<z<4$, we assume for the forecast that Spec-S5 uses only the $u$- and $g$-dropout samples.
This defines an expected LBG sample covering an area of $11,000\,{\rm deg}^{2}$, see for example Table~2 of Ref.~\cite{2025arXiv250307923B}.
The main properties of Spec-S5 are summarized in the last section of \autoref{tab:surveys}.

Throughout the forecast, we standardize the redshift bins for the Stage-V LBG samples.
For LBG $u$-dropouts, we consider a flat distribution over $2.5<z<3.5$.
For $g$-dropouts, we use $3.5<z<4.5$, and for $r$-dropouts, $4.5<z<5.5$.
We also standardize the expected linear galaxy bias following the relation \cite{2019JCAP...10..015W}
\begin{equation}
    b(z,m)=A(m)(1+z)+B(m)(1+z)^{2},
\end{equation}
where $A(m)=-0.98(m-25)+0.11$ and $B(m)=0.12(m-25)+0.17$, with $m$ being the magnitude of the galaxies within each sample.
Although this relation predicts exact bias values, we use it only as an estimate and assign the same fiducial bias to each dropout population across the different surveys.
The actual galaxy bias values used for the forecast are presented in the fifth column of \autoref{tab:surveys}.
This choice provides a fair comparison between surveys, so that differences in the forecast are mainly driven by survey properties such as number density, area, and redshift coverage.

\section{PNG constraints from DESI Imaging data}\label{sec:delta_desi}

Signatures of $\fnl^{\rm loc}$ could contain hidden information about massive fields during inflation through $\Delta$.
To explore this possibility in a real data set, we use the potential local PNG signals found in Rezaie24 as a starting point.
We first confirm that our pipeline does recover the reported constraints on $\fnl^{\rm loc}$ from Rezaie24.
For this purpose, we use the ACF pipeline described in \autoref{sec:inference}, with the parameter space $\Theta=\{\fnl^{\rm loc},b\}$.
In this analysis, following Rezaie24, we sample over the growth-normalized bias $b$, related to the physical galaxy bias by $b(z_{\rm eff}) = b/D(z_{\rm eff})$.
We obtain constraints on $\fnl^{\rm loc}$ for each set of weight maps, as well as for the galaxy sample without observational systematics decontamination.
We also repeat the analysis for different angular scale cuts in order to test the stability of the local PNG constraints and to study how the inclusion of smaller angular scales affects the beyond-local PNG fit.

\begin{figure}
\centering
\includegraphics[width=\columnwidth]{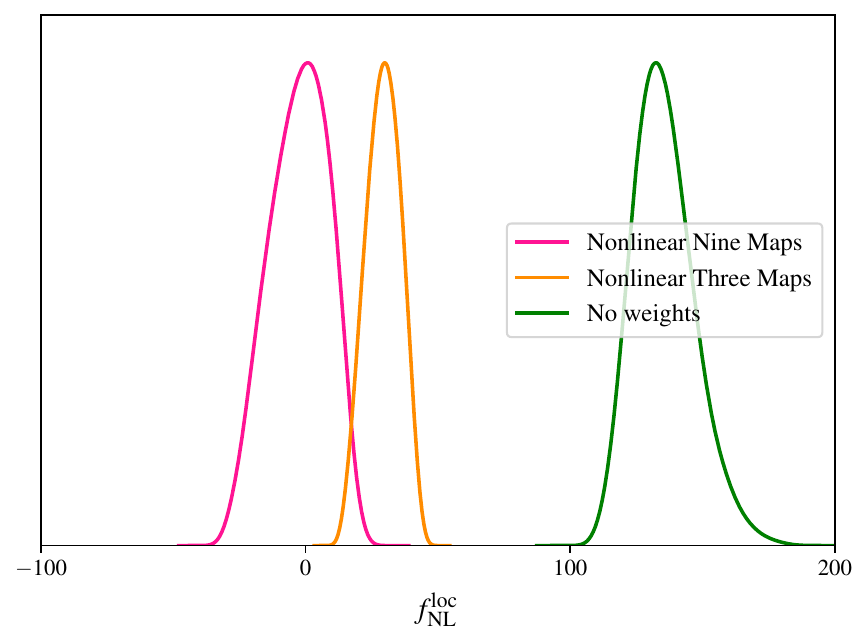}
\caption{Marginalized posterior distributions for the local PNG amplitude $\fnl^{\rm loc}$ for the DESI Imaging LRG sample using the angular correlation function for $0.5\deg < \theta < 20\deg$. The pink posterior corresponds to the decontaminated case using \textit{Nonlinear Nine Maps}, while the orange posterior corresponds to using \textit{Nonlinear Three Maps}. The green posterior shows the unweighted sample without observational systematics decontamination. These constraints are not calibrated for overcorrection.
The results are very similar to the angular power spectrum analysis of Ref.~\cite[see their Fig.~14]{2024MNRAS.532.1902R}}
\label{fig:fnlloc_weights}
\end{figure}

The results of the local PNG analysis are shown in \autoref{fig:fnlloc_weights}, where we show the 1D marginalized posterior distribution for $\fnl^{\rm loc}$ for different observational systematics mitigation approaches.
As discussed in \autoref{sec:desi_data}, the application of these weights can suppress the true large-scale clustering signal through overcorrection, leading to a biased $\fnl^{\rm loc}$ measurement. 
While overcorrection can be calibrated using simulations with injected local PNG signals \cite{2024MNRAS.532.1902R}, we omit this calibration in this work. 
The aim of this analysis is not to present a new measurement of $\fnl^{\rm loc}$, but to use these uncalibrated configurations to explore exactly how apparent local PNG signals, due in this case to residual systematics, project into the beyond-local parameter space.
The resulting local PNG constraints are presented in the second column of \autoref{tab:desi_constraints_scale_cuts}.

\begin{table*}

\centering
\caption{Constraints from the DESI Imaging ACF analysis for different angular scale cuts and observational systematics treatments. The local PNG constraints are obtained with $\Theta=\{\fnl^{\rm loc},b\}$, while the beyond-local PNG constraints use $\Theta=\{\fnld,\Delta,b\}$.
The column $P(\fnld>0)$ gives the marginalised posterior mass for a positive beyond-local PNG amplitude and is used only as a diagnostic of the shape of the posterior, not as a detection significance.
Entries in the $\Delta$ column written as lower limits correspond to one-sided $95\%$ marginalised credible limits.
Bold row corresponds to the configuration that gives a preference for $\Delta$.}
\label{tab:desi_constraints_scale_cuts}
\setlength{\tabcolsep}{6pt}
\renewcommand{\arraystretch}{1.4}
\begin{tabular}{lcccc}
\hline\hline
Weights & $\fnl^{\rm loc}$ & $P(\fnld>0)$ & $\fnld$ & $\Delta$ \\
\hline\hline

\multicolumn{5}{l}{$1\deg < \theta < 20\deg$} \\
\hline
Nonlinear Nine Maps  & $-5^{+15}_{-13}$ & $0.772$ & $1.87^{+5.28}_{-2.12}\times 10^{4}$ & $>0.88$ \\
Nonlinear Three Maps & $22^{+10}_{-8}$ & $0.981$ & $2.08^{+4.06}_{-1.70}\times 10^{4}$ & $>0.59$ \\
No weights           & $100^{+10}_{-20}$ & $1.000$ & $5.56^{+5.57}_{-3.00}\times 10^{2}$ & $0.24^{+0.09}_{-0.10}$ \\

\hline
\multicolumn{5}{l}{$1\deg < \theta < 25\deg$} \\
\hline
Nonlinear Nine Maps  & $-7\pm 14$ & $0.741$ & $1.51^{+5.24}_{-1.89}\times 10^{4}$ & $>0.87$ \\
Nonlinear Three Maps & $20^{+10}_{-7}$ & $0.985$ & $1.94^{+3.99}_{-1.60}\times 10^{4}$ & $>0.54$ \\
No weights           & $95^{+10}_{-20}$ & $1.000$ & $4.79^{+4.48}_{-2.40}\times 10^{2}$ & $0.22^{+0.08}_{-0.09}$ \\

\hline
\multicolumn{5}{l}{$0.5\deg < \theta < 25\deg$} \\
\hline
Nonlinear Nine Maps  & $-3^{+16}_{-14}$ & $0.409$ & $-0.44^{+1.68}_{-2.17}\times 10^{4}$ & $>0.83$ \\
\textbf{Nonlinear Three Maps} & $\bm{27^{+10}_{-9}}$ & $\bm{0.945}$ & $\bm{5.12^{+6.13}_{-3.62}\times 10^{3}}$ & $\bm{0.91^{+0.25}_{-0.19}}$ \\
No weights           & $136^{+11}_{-13}$ & $1.000$ & $6.06^{+4.15}_{-2.64}\times 10^{2}$ & $0.22^{+0.07}_{-0.08}$ \\

\hline
\end{tabular}
\end{table*}

We first consider the \textit{Nonlinear Nine Maps} case, which corresponds to the main result reported in the final version of Rezaie24.
For this weight map, we find $\fnl^{\rm loc}=-5^{+15}_{-13}$, $-7\pm14$, and $-3^{+16}_{-14}$ for the three scale cuts considered, all consistent with no significant evidence for local PNG.
This case provides the reference result in which the decontaminated data do not show a local PNG signal.

A more interesting case is obtained with the \textit{Nonlinear Three Maps} weights.
For this decontamination method, we find $\fnl^{\rm loc}=22^{+10}_{-8}$, $20^{+10}_{-7}$, and $27^{+10}_{-9}$ for the three scale cuts.
This indicates a hint of local PNG that survives observational systematics decontamination, likely indicating remaining systematics, according to Rezaie24.
This hint motivates testing to see whether it can be interpreted as a signature of $\Delta$ in real data.\footnote{Note that this decontamination scheme was the main reported result in the first version of Rezaie24. This motivated our exploration of PNG beyond the local type.}
The comparison between $1\deg<\theta<20\deg$ and $1\deg<\theta<25\deg$ shows that extending the maximum angular separation does not qualitatively change this local PNG hint.

Finally, we also consider the sample without weight correction.
In this case, we find a much stronger apparent signal, $\fnl^{\rm loc}=100^{+10}_{-20}$, $95^{+10}_{-20}$, and $136^{+11}_{-13}$, depending on the scale configuration.
Since no observational systematics decontamination is applied, this signal is interpreted as spurious.
This confirms the expected result that observational systematics can induce apparent local PNG signals if they are not properly removed.

The comparison between scale cuts shows that the qualitative local PNG conclusions are stable across the angular ranges considered. The \textit{Nonlinear Nine Maps} case remains consistent with no local PNG signal, the \textit{Nonlinear Three Maps} case shows a moderate and stable hint, and the unweighted sample shows a strong apparent signal for all scale cuts. The variations in the central values are modest compared with the size of the corresponding $\fnl^{\rm loc}$ signals, especially when compared with the difference between the decontaminated and unweighted cases.

These results provide the first application to real data of the ACF pipeline presented in Riquelme23, and the tightest constraints on local PNG obtained so far using the ACF.
The local PNG constraints are in remarkable agreement with the DESI Imaging analysis of Rezaie24 in harmonic space for all three mitigation strategies.
This agreement provides a validation of the ACF pipeline on real data.

\begin{figure}
\centering
\includegraphics[width=\columnwidth]{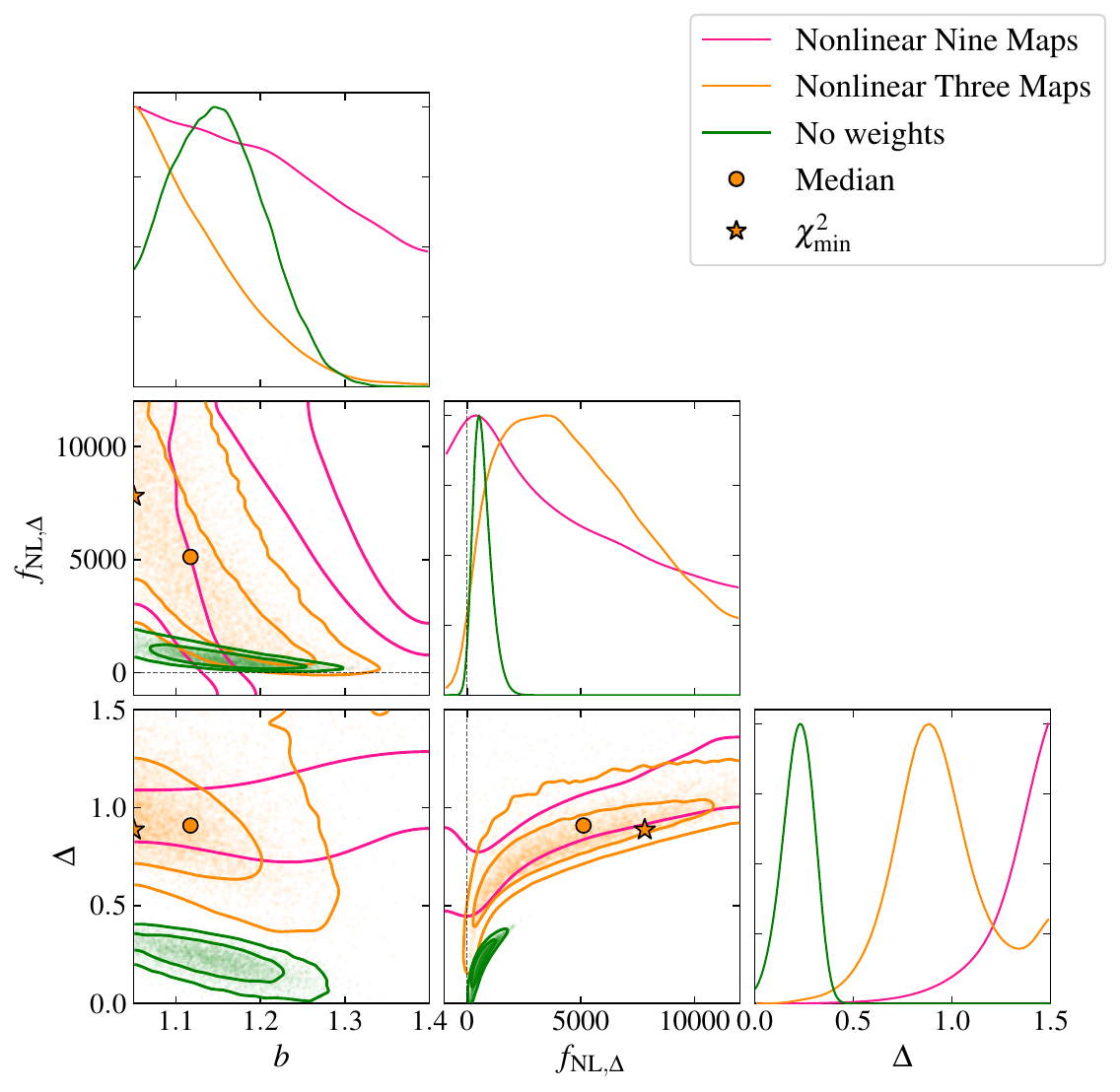}
\caption{Marginalized contours for the linear bias $b$, the beyond-local PNG amplitude $\fnld$, and the scaling exponent $\Delta$ for the DESI Imaging LRG sample, measured over $0.5\deg< \theta < 25\deg$. The pink contours correspond to the decontaminated sample using \textit{Nonlinear Nine Maps}, while the orange contours correspond to \textit{Nonlinear Three Maps}. The green contours show the unweighted sample, without observational systematics decontamination.
For the \textit{Nonlinear Three Maps} case, the markers show the 1D median and the best-fit point from $\chi^2$ minimization.
The $\fnld$ range has been zoomed in to highlight the relative differences between the contours. The constraints corresponding to these contours are presented in the last section of \autoref{tab:desi_constraints_scale_cuts}. Similarly to the local PNG constraints, the beyond-local PNG constraints are not calibrated for overcorrection and are therefore subject to mitigation systematics.}
\label{fig:weight_comp}
\end{figure}

After confirming that the ACF pipeline reproduces the DESI Imaging LRG constraints for local PNG, we extend parameter space to now include the effect of massive fields, $\Theta=\{\fnld,\Delta,b\}$.
\autoref{fig:weight_comp} shows the results for
$0.5\deg<\theta<25\deg$, while the last three columns of
\autoref{tab:desi_constraints_scale_cuts} summarize the constraints for $\fnld$ and $\Delta$ for all the scale cuts.
As noted in \autoref{sec:desi_data}, $\theta_{\rm min}=0.5\deg$ approximately marks the boundary of the linear regime. 
The validity of pushing to this scale is empirically supported by the stability of our local PNG constraints across the different angular cuts (see the second column in \autoref{tab:desi_constraints_scale_cuts}).

The pink contours correspond to the \textit{Nonlinear Nine Maps} decontamination case and show the expected divergent behavior between $\Delta$ and $\fnld$, as described by \autoref{eq:scale_dep}, when no significant local PNG signal is detected.
In this case, following the reported constraints in \autoref{tab:desi_constraints_scale_cuts}, since the PNG amplitude is consistent with zero, the scale dependence is unconstrained, creating a funnel-shaped degeneracy between $\fnld$ and $\Delta$.
Because the degeneracy extends toward large $\Delta$, the marginalized posterior is further shaped by the prior volume, causing the posterior mass to accumulate near the upper prior boundary and yielding the reported one-sided lower limit.
This behavior is stable across the different scale cuts and reflects the persistence of the likelihood degeneracy when the local PNG amplitude is consistent with zero.

For \textit{Nonlinear Three Maps}, shown by the orange contours, the $\theta_{\rm min}=1\deg$ fits retain
the high-$\Delta$ tail and yield only one-sided limits. Extending
$\theta_{\rm max}$ from $20\deg$ to $25\deg$ does not improve the
constraint.
These two results are consistent with each other and show that extending the maximum angular separation alone does not improve the constraints on $\Delta$.

When the smaller angular scales are included, the posteriors become better controlled and give $\fnld=5.12^{+6.13}_{-3.62}\times10^{3}$ and $\Delta=0.91^{+0.25}_{-0.19}$, as shown by the orange contour in \autoref{fig:weight_comp}.
This suggests that the additional small-scale information helps to control the divergent direction between $\fnld$ and $\Delta$, allowing the apparent local PNG hint to be mapped into a preference for $\Delta$.
This behavior is expected given the shape of the scale-dependent bias, $\propto\fnld k^{\Delta -2}$.
For larger values of $\Delta$, the PNG contribution to the galaxy bias varies less across scales than in the local PNG case, reducing the large-scale enhancement relative to smaller scales.
The resulting wider range of scales helps distinguish a change in $\Delta$ from a change in $\fnld$, breaking the degeneracy between the two parameters.

For this scale cut, the marginalized posterior assigns most of its probability to positive values of the PNG amplitude, with $P(\fnld>0)=0.945$.
This posterior probability is influenced by prior boundaries and the associated prior-volume effects acting on the remaining parameter degeneracies.
The preference around $\Delta\sim0.9$ is associated with the positive-$\fnld$ region traced by the main orange contour.
The marked points show that the 1D median and best-fit values of $\Delta$ are in close agreement, whereas the corresponding values of $\fnld$ differ due to the non-Gaussian shape of its marginalized posterior.

The global minimum of the beyond-local PNG fit over the \textit{Nonlinear Three Maps} occurs at $(\fnld,\Delta,b)=(7.81\times10^{3},0.89,1.05)$. In the local limit ($\Delta=0$), the best fit is $(\fnl^{\rm loc},b)=(30.3,1.28)$, yielding $\Delta\chi^2=3.95$.
This corresponds to an approximate $2.3\sigma$ preference for the beyond-local PNG model over the local PNG model.
The previous significance is computed using Chernoff's theorem \cite{10.1214/aoms/1177728725}, which applies when the null local PNG hypothesis lies on the boundary of the parameter space of the alternative beyond-local model at $\Delta=0$.
For reference, the Gaussian fit, $f_{\rm NL}=0$, gives $b=1.30$, corresponding to a $2.65\sigma$ preference for the local PNG model ($\Delta\chi^2=7.04$) under Wilks' theorem \cite{10.1214/aoms/1177732360}.
These comparisons indicate that the previously identified hint for local PNG is accompanied by an additional preference for a beyond-local PNG model.

The green contours correspond to the sample without decontamination, where a spurious $\fnl^{\rm loc}$ signal was found.
The unweighted sample also yields closed constraints,
$\Delta\simeq0.22$--$0.24$, showing that a strong spurious local PNG signal
can generate an apparently constrained beyond-local posterior, even when the origin of the signal is clearly not cosmological.

Since the local PNG signal in DESI Imaging data is likely affected by residual observational systematics, the corresponding preference for non-zero $\Delta$ should not be interpreted as evidence for extra massive fields.
Instead, the beyond-local PNG fit identifies the region of $\Delta$ that would be preferred if the apparent local PNG signal were cosmological.
This first application of the beyond-local PNG model to a data set with an apparent local-PNG hint illustrates how such a signal projects into the $(\fnld,\Delta)$ posterior and highlights the impact of residual degeneracies and observational systematics on future searches for massive-field signatures.

An additional limitation is the uncalibrated overcorrection induced by the systematics weights. Calibrating this effect for beyond-local PNG would require simulations with injected $\Delta$ signatures, which are not currently available. A calibration of overcorrection would become necessary only if a stable preference for $\Delta$ were found across different mitigation schemes.

\section{Forecast for Stage-V surveys}\label{sec:delta_stagev}

Considering the configuration of the different Stage-V surveys introduced in \autoref{sec:stagev_data}, here we study the constraints that they will be able to put on the parameter $\Delta$. 
Before that, we first establish the baseline sensitivity of each survey to local PNG.
The second column of \autoref{tab:results} summarizes the forecasted $\fnl^{\rm loc}$ constraints for a baseline model with fiducial $\fnl^{\rm loc}=4$ and marginalizing over galaxy bias. We recover the expected uncertainties (second column) for $\fnl^{\rm loc}$ presented in Ref.~\cite{2023MNRAS.521.3648D} for MUST. We also find that WST improves the constraints by approximately a factor of 2 with respect to MUST, consistent with Ref.~\cite{2024arXiv240305398M}. For Spec-S5, we obtain forecasted constraints compatible with those presented in Ref.~\cite{2025arXiv250307923B}.
\begin{table*}
\centering
\small
\setlength{\tabcolsep}{6pt}
\renewcommand{\arraystretch}{1.4}
\caption{Summary of the Stage-V forecasts. The constraints on $\Delta$ are obtained from theory data vectors with $\Delta^{\rm fid}=0$, while the constraints on $\fnl^{\rm loc}$ assume a fiducial value of $\fnl^{\rm loc}=4$. 
We report results for both the free-bias case, marginalizing over galaxy
bias, and the fixed-bias case.}
\begin{tabular}{lccccc}
\hline \hline
Survey & $\sigma(\fnl^{\rm loc})$ & $\sigma^{\rm free}(\Delta),\ \fnldfid=4$ & $\sigma^{\rm fixed}(\Delta),\ \fnldfid=4$ & $\sigma^{\rm free}(\Delta),\ \fnldfid=18$ & $\sigma^{\rm fixed}(\Delta),\ \fnldfid=18$ \\
\hline \hline
WST    & 0.58 & 0.17 & 0.072 & 0.061 & 0.024 \\
MUST   & 0.84 & 0.34 & 0.114 & 0.088 & 0.036 \\
Spec-S5 & 1.06 & 0.50 & 0.135 & 0.109 & 0.042 \\
\hline
\end{tabular}
\label{tab:results}
\end{table*}

\subsection{Around the local PNG limit, $\Delta^{\rm fid}=0$}

Once we have established that our pipeline recovers the expected forecasted constraints for local PNG, we move on to explore the constraining power on beyond-local shapes via $\Delta$. We start by setting our theory vector with $\Delta^{\rm fid}=0$ and obtain constraints for different values of $\fnldfid$ for WST, MUST and Spec-S5.
The choice $\Delta^{\rm fid}=0$ corresponds to a fiducial local PNG signal. This allows us to quantify how well a survey that detects local PNG could constrain departures from the local limit.

We run the pipeline as described in \autoref{sec:inference}. 
From the sampled posterior distribution, we consider two cases: one where the galaxy bias is marginalized over (free bias), and one where it is fixed to its fiducial value (fixed bias). 
In both cases, $\fnld$ is marginalized over. We then use the resulting posterior distribution of $\Delta$ to extract the forecasted $68\%$ C.L., $\sigma(\Delta)$.
\begin{figure}
    \centering
    \includegraphics[width=\columnwidth]{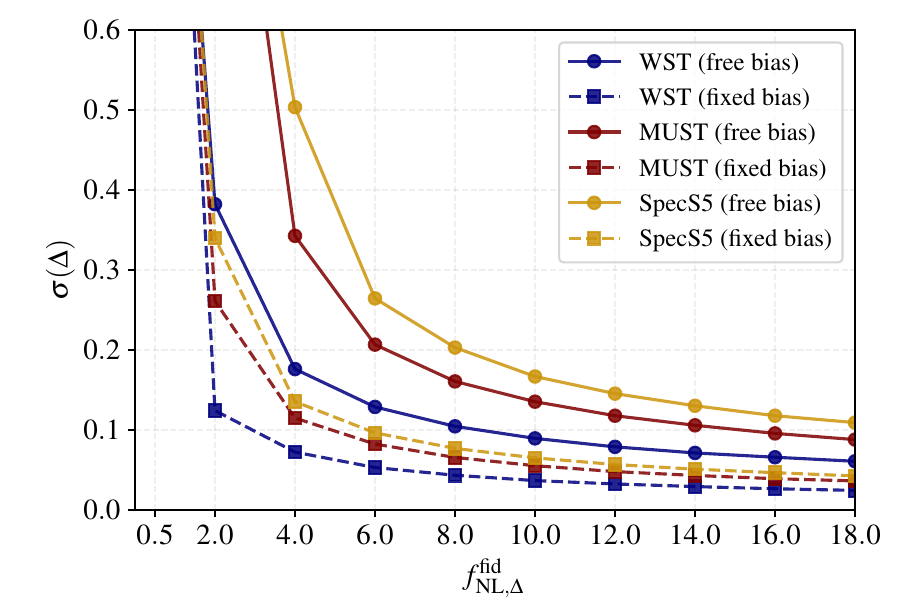}
    
    \vspace{0.4cm} 
    
    \includegraphics[width=\columnwidth]{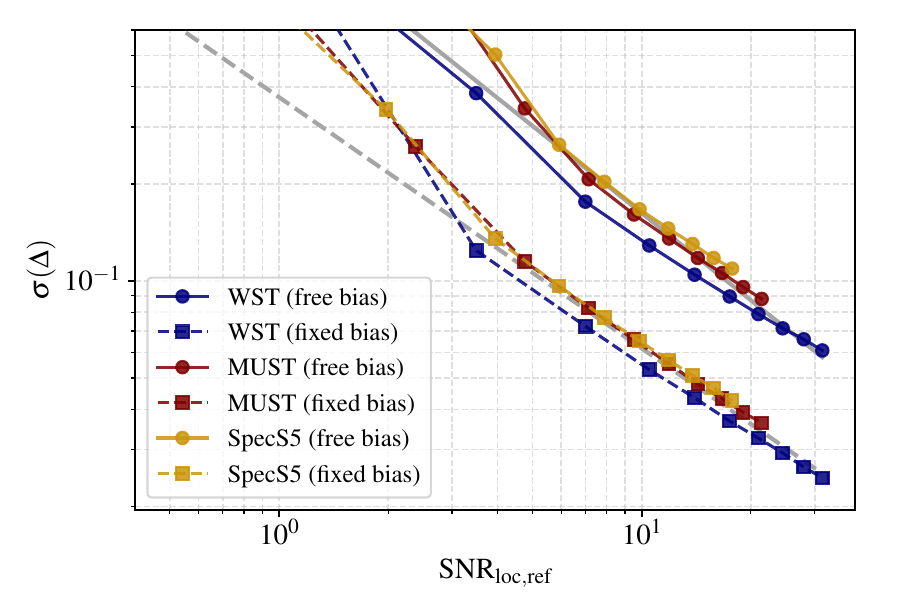}
    
    \caption{\textbf{Top panel:} Marginalized $1\sigma$ forecasted constraints as a function of $\fnldfid$ for each Stage-V survey. Solid lines correspond to the free-bias case, while dashed lines correspond to the fixed-bias case. \textbf{Bottom panel:} Marginalized constraints as a function of $\text{SNR}_{\rm loc, ref}$ defined by \autoref{eq:snrref}. Grey lines show the fitted power law relations.}
    \label{fig:sigma_delta_fnl}
\end{figure}

The top panel of \autoref{fig:sigma_delta_fnl} shows the forecasted constraints on $\Delta$ as a function of $\fnldfid$ for each survey.
The constraints improve as $\fnldfid$ increases because a larger PNG amplitude increases the sensitivity to its scale dependence.
For $\fnldfid<4$, however, the $(\fnld,\Delta)$ posteriors become funnel-shaped and prior-dominated.
The resulting marginalized constraints on $\Delta$ are affected by prior-volume effects and should not be interpreted as forecast constraints since they do not recover the fiducial values.

The last four columns of \autoref{tab:results} present $\sigma(\Delta)$ for selected values of $\fnldfid$ in the free- and fixed-bias cases.
Fixing the galaxy bias reduces the marginalized uncertainty on $\Delta$ by approximately $60\%$ across the Stage-V surveys and PNG amplitudes considered. 
This shows that the degeneracy between galaxy bias and $\fnld$ is a major limitation and that external bias information could improve the detectability of massive-field signatures.
For an amplitude consistent with current Planck constraints \cite{2020A&A...641A...9P}, $\fnldfid=\fnl^{\rm loc}\simeq4$ around the local limit, we find $\sigma(\Delta)\simeq0.17$--$0.50$, depending on the survey.
For $\fnldfid=18$, this improves to $\sigma(\Delta)\simeq0.06$--$0.11$ in the free-bias case.

The top panel of \autoref{fig:sigma_delta_fnl} suggests a power-law scaling of the constraints on $\Delta$ as a function of $\fnldfid$ across the different surveys. An interesting behavior appears when, instead, we express the results in terms of the fiducial beyond-local PNG amplitude normalized by the local PNG sensitivity of each survey. We define
\begin{equation}\label{eq:snrref}
    \text{SNR}_{\rm loc,ref} \equiv
    \frac{\fnldfid}{\sigma(\fnl^{\rm loc})},
\end{equation}
where $\sigma(\fnl^{\rm loc})$ is the expected constraint on $\fnl^{\rm loc}$ for each survey in the local PNG template, without including $\Delta$.
We interpret this quantity as a reference local PNG significance. It tells us how significant a local PNG signal would be in the same survey if its amplitude were set to $\fnldfid$.
In this section, where $\Delta^{\rm fid}=0$, this is equivalent to the usual local PNG significance. For $\Delta^{\rm fid}\neq0$, it should instead be understood as a reference sensitivity, not as the true detection significance of the non-local signal.

The bottom panel of \autoref{fig:sigma_delta_fnl} shows the forecasted constraints on $\Delta$ as a function of $\text{SNR}_{\rm loc,ref}$. 
The figure reveals a transition in the behavior of the constraints. At low significance, $\text{SNR}_{\rm loc,ref}<4$, as mentioned before, surveys do not have enough constraining power to resolve the shape of the posterior. As a result, the marginalized constraint remains affected by the divergence region induced by the degeneracy between $\fnld$ and $\Delta$. 
Once the reference significance exceeds this threshold, the posterior detaches from this region, and the contours recover an approximately elliptical shape.
In this regime, the results from the different surveys collapse onto a common power law, shown in solid gray for the free-bias case and in dashed gray for the fixed-bias case.
This indicates that the improvement in the constraints on $\Delta$ is not set by the absolute value of $\fnldfid$ alone. Instead, the relevant quantity is how large this fiducial amplitude is compared to the local PNG uncertainty of each survey.
The relation holds across surveys with different redshift ranges, areas, and number densities, suggesting that the dominant survey dependence is captured by $\sigma(\fnl^{\rm loc})$ for the forecast setup considered here.

For the free-bias case, the common trend is well described by the empirical relation
\begin{equation}
    \sigma(\Delta) \simeq 1.27 \, {\rm SNR}_{\rm loc,ref}^{-0.90}\, ,
\end{equation}
with coefficient of determination, $R^2 = 0.96$. This near-inverse scaling is expected near the local limit, since the response of the scale-dependent bias to $\Delta$ is proportional to the PNG amplitude. In this regime, the fiducial signal is close to the local PNG shape, so $\sigma(\fnl^{\rm loc})$ provides a useful formalization for the baseline PNG sensitivity of each survey. The fitted slope is close to $-1$ across the different cases, indicating that, near the local limit, the constraining power on $\Delta$ is mainly controlled by the fiducial beyond-local PNG amplitude measured relative to the local PNG sensitivity. This also provides a useful way to account for small differences between forecast setups, since changes in the assumed survey properties enter mainly through the baseline local PNG sensitivity.

\subsection{Beyond $\Delta^{\rm fid}=0$}

Having established the constraining power around the local limit, we now extend the forecast to non-zero fiducial values of $\Delta$. 
The goal of this section is to quantify how the constraints on both $\fnld$ and $\Delta$ change as the fiducial signal moves away from the local PNG shape.
This is important because larger values of $\Delta$ weaken the scale dependence of the PNG contribution and increase its degeneracy with galaxy bias.
Motivated by this degeneracy, we compare free- and fixed-bias configurations.
We take the free-bias case as the baseline forecast, while the fixed-bias case is shown as an optimistic comparison.

\begin{figure}
    \centering
    \includegraphics[width=\columnwidth]{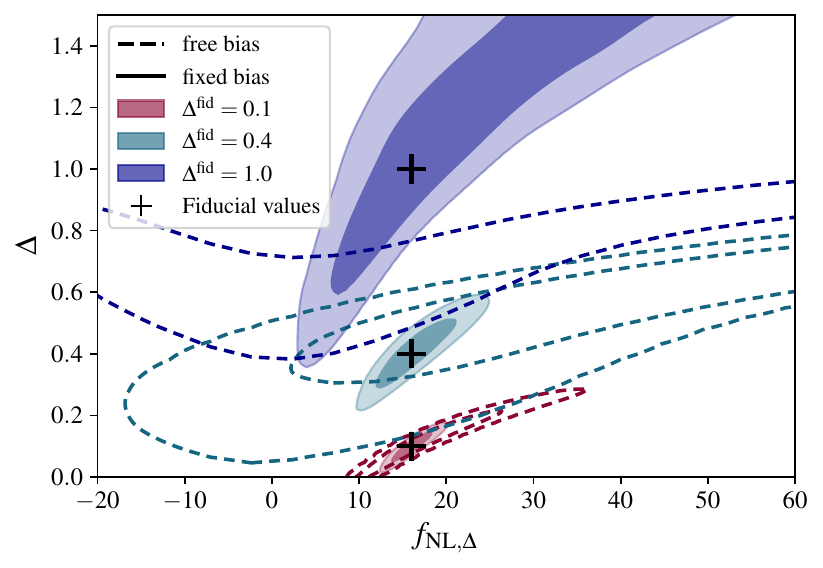}
    \caption{Forecasted free- and fixed-bias contours for WST, assuming $\fnldfid=16$ and $\Delta^{\rm fid}=\{0.1,0.4, 1.0\}$, shown by the crosses. Dashed contours correspond to the free-bias case, while solid contours correspond to the fixed-bias case. The figure shows that the constraints degrade for larger values of $\Delta^{\rm fid}$ and highlights the importance of galaxy bias degeneracies.}
    \label{fig:delta_nozero}
\end{figure}

To explicitly see how the constraining power degrades for larger values of $\Delta^{\rm fid}$, we show in \autoref{fig:delta_nozero} the marginalized contours for $\fnld$ and $\Delta$.
For this test, we assume a fixed PNG amplitude of $\fnldfid=16$ and evaluate $\Delta^{\rm fid}=\{0.1,0.4, 1.0\}$, which correspond to extra fields with masses of $m \approx 0.5H$, $m \approx 1.0H$ and $m=\sqrt{2}H$, respectively.
For simplicity, this analysis is performed using the expected properties of the WST survey.
We use WST as a representative case because the previous subsection showed that much of the survey dependence is captured by the baseline local PNG sensitivity.

The contours show the impact of galaxy bias degeneracies. In the free-bias case, the 2D posteriors broaden significantly compared to the fixed-bias case.
Furthermore, the figure confirms the degradation of the constraining power as $\Delta^{\rm fid}$ increases.
This behavior is expected from \autoref{eq:scale_dep}, since larger values of $\Delta$ weaken the scale-dependent signal.
As $\Delta^{\rm fid}$ increases, the uncertainty on $\fnld$ grows together with $\sigma(\Delta)$, showing that amplitude and shape become harder to recover simultaneously.

For the free-bias WST case with $\fnldfid=16$ and $\Delta^{\rm fid}=0.1$, presented in dashed red contours in \autoref{fig:delta_nozero}, we recover the fiducial values of both $\fnld$ and $\Delta$ at $1\sigma$, with $\Delta = 0.13^{+0.08}_{-0.07}$ and $\fnld = 17.9^{+6.9}_{-4.7}$.
For $\Delta^{\rm fid}=0.4$, shown in dashed green contours, the uncertainties become larger.
For example, the measurement gives $\fnld=58^{+90}_{-40}$, so the fiducial value remains within the $1\sigma$ uncertainty, although the PNG amplitude is much less significant.
At the same time, we almost recover the fiducial value of $\Delta$, with $\Delta=0.68^{+0.18}_{-0.25}$.
Finally, for $\Delta^{\rm fid}=1.0$, in the dashed dark blue contours, the posterior reaches the upper edge of the $\Delta$ prior.
In the free-bias case, we find $\fnld=6.1^{+66.7}_{-59.0}$ and report only the one-sided $95\%$ marginalized lower limit, $\Delta>0.60$.
This constraint on $\Delta$ should not be interpreted as a closed measurement, since the posterior extends to the upper boundary of the $\Delta$ prior.

This shows that, for large values of $\Delta^{\rm fid}$, the scale dependence becomes too weak to recover both $\Delta$ and $\fnld$ with the same precision as in the lower $\Delta^{\rm fid}$ cases.
This behavior is similar to what we find in the DESI data: when moving from the local PNG to the beyond-local PNG, the uncertainty on the amplitude increases substantially, while the posterior can still show a preference for non-zero $\Delta$.

When the bias values are fixed, the posteriors become closer to Gaussian and the uncertainties decrease.
The fixed-bias case therefore provides an optimistic estimate of the intrinsic scaling of $\sigma(\Delta)$ with the PNG amplitude.

We next test whether the scaling with local PNG sensitivity also holds for non-zero $\Delta^{\rm fid}$.
To do this, we repeat the forecast for $\Delta^{\rm fid}=0.1$ to $1.0$ and evaluate the uncertainty on $\Delta$ as a function of $\text{SNR}_{\rm loc,ref}$.
Since the low significance regime can still be affected by projection effects from the funnel-shaped degeneracy, we impose an SNR cutoff for each fiducial value of $\Delta$ before fitting the scaling relation.

\begin{figure}
    \centering
    \includegraphics[width=\columnwidth]{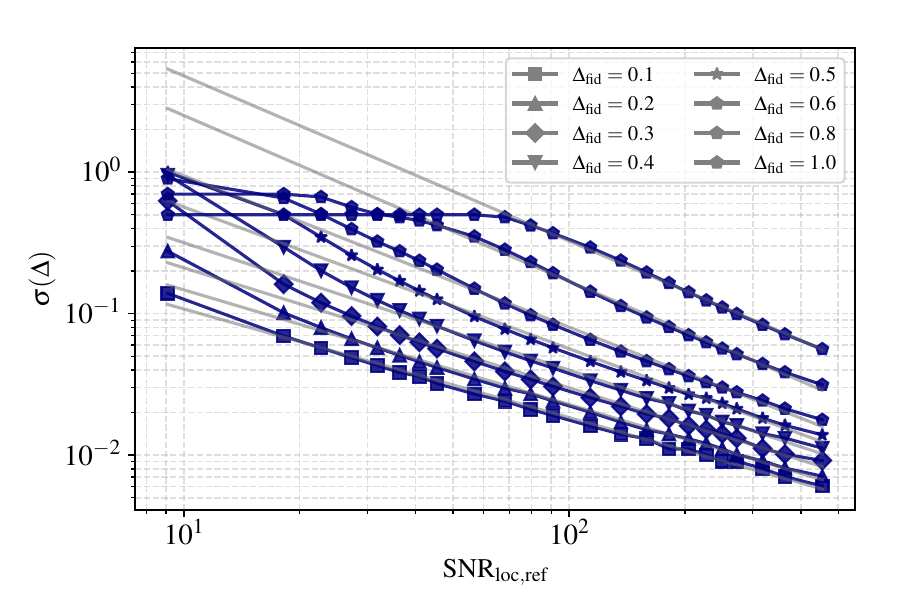}
    
    \vspace{0.02cm}
    
    \includegraphics[width=\columnwidth]{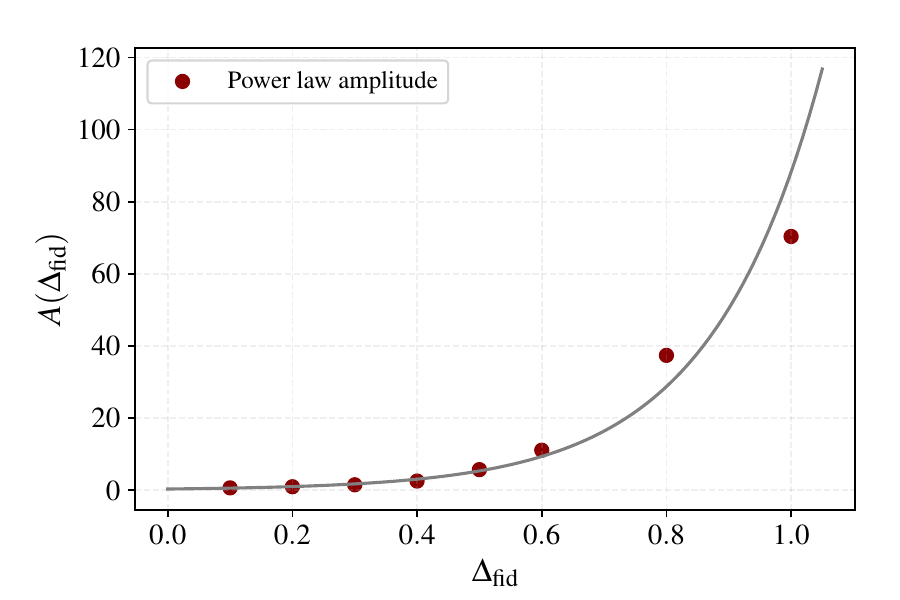}
    
    \caption{\textbf{Top panel:} Forecasted constraints on $\Delta$ as a function of $\text{SNR}_{\rm loc,ref}$ for different fiducial values of $\Delta$. Gray lines correspond to the power law fits for the high SNR regime. \textbf{Bottom panel:} Amplitude $A(\Delta^{\rm fid})$ of the fitted relation as a function of $\Delta^{\rm fid}$. The solid line shows the exponential fit to the measured amplitudes.}
    \label{fig:deltafid_snr_scaling}
\end{figure}

As shown in \autoref{fig:deltafid_snr_scaling}, after removing the low significance points affected by projection effects, the constraints are well described by the empirical relation
\begin{equation}
    \sigma(\Delta) \simeq A(\Delta^{\rm fid}) \, 
    \text{SNR}_{\rm loc,ref}^{-m(\Delta^{\rm fid})}.
\end{equation}
The shift of the fitted relation with $\Delta^{\rm fid}$ shows that larger departures from the local limit require a smaller value of $\sigma(\fnl^{\rm loc})$, or a larger fiducial amplitude $\fnldfid$, to reach the same uncertainty on $\Delta$. While the fitted slopes are close to the local limit shapes from the previous section, the power law steepens as $\Delta^{\rm fid}$ increases.
This behavior can be described by
\begin{equation}
   m(\Delta^{\rm fid})\simeq 0.72 + 0.50 \Delta^{\rm fid},
\end{equation}
with a coefficient of determination $R^2 = 0.94$. 
The normalization of the relation also grows rapidly with $\Delta^{\rm fid}$. Using the fixed-bias configuration to isolate the relation between $\fnld$ and $\Delta$, we find
\begin{equation}
    A(\Delta^{\rm fid}) \simeq 0.32 \, e^{5.60 \Delta^{\rm fid}},
\end{equation}
with $R^2=0.989$. This relation for the fixed-bias case should be interpreted as a best-case scenario of the intrinsic scaling between $\fnld$ and $\Delta$, rather than as the baseline forecast.

For a target uncertainty $\sigma_{\rm target}(\Delta)$, the same relation can be inverted to estimate the local PNG sensitivity required for a model with predicted amplitude $\fnldfid$,
\begin{equation}\label{eq:sigfnlreq}
    \sigma_{\rm req}(\fnl^{\rm loc})
    \simeq \fnldfid
    \left[ \frac{\sigma_{\rm target}(\Delta)}{A(\Delta)}
    \right]^{1/m(\Delta)}.
\end{equation}
This expression provides an estimate for translating a prediction for $\fnldfid$ and $\Delta$ into the local PNG constraining power needed to measure $\Delta$.
In Table~\ref{tab:sigma_req_examples}, we show some representative examples using the same values of $\fnldfid$ and $\Delta^{\rm fid}$ considered above, assuming a target significance of $3\sigma$, i.e. $\sigma_{\rm target}(\Delta)=\Delta/3$.
\begin{table}
\centering
\caption{Best-case fixed-bias estimate of the local PNG uncertainty required to obtain a $3\sigma$ measurement of $\Delta$ for representative model predictions.}
\label{tab:sigma_req_examples}
\small
\setlength{\tabcolsep}{6pt}
\renewcommand{\arraystretch}{1.4}
\begin{tabular}{cccc}
\hline \hline
$\Delta$ & $\fnldfid$ & $\sigma_{\rm target}(\Delta)$ & $\sigma_{\rm req}(\fnl^{\rm loc})$ \\
\hline \hline
$0.1$ & $16$ & $0.033$ & $0.41$ \\
$0.4$ & $16$ & $0.133$ & $0.54$ \\
$1.0$ & $16$ & $0.333$ & $0.17$ \\
\hline
\end{tabular}
\end{table}
The table is included to illustrate how the fitted scaling can be translated into a survey requirement. Since it is based on the fixed-bias calibration, the values should be interpreted as optimistic requirements; marginalizing over galaxy bias would require a smaller tolerated value of $\sigma(\fnl^{\rm loc})$ to reach the same target uncertainty on $\Delta$.
Since \autoref{eq:sigfnlreq} scales linearly with $\fnldfid$, the required sensitivity for other amplitudes can be obtained by simple rescaling.

Finally, these results show that Stage-V surveys can test signatures of massive fields only if the fiducial beyond-local PNG amplitude is sufficiently large relative to the local PNG sensitivity of the survey. Around the local limit, the expected constraints approximately follow the scaling $\sigma(\Delta)\propto \text{SNR}_{\rm loc,ref}^{-1}$. For non-zero $\Delta^{\rm fid}$, the required sensitivity increases because the scale-dependent signal becomes weaker and more degenerate with the PNG amplitude and the galaxy bias parameters.

\section{Conclusions}\label{sec:Conclusions}

We have presented an analysis of signatures of massive fields during inflation using the scale-dependent bias in galaxy clustering.
These signatures are parametrized by the amplitude $\fnld$ and the scaling parameter $\Delta$, which controls the departure from the local PNG limit (see \autoref{eq:scale_dep}).
In this model, the local case is recovered for $\Delta=0$, while non-zero values of $\Delta$ weaken the scale dependence of the PNG contribution to the galaxy bias.
This makes $\Delta$ difficult to constrain, especially when the PNG amplitude is small or compatible with zero.

We extended the angular correlation function pipeline previously used in Riquelme23 to constrain local PNG to include the effect of $\Delta$.
We first applied this model to the DESI Legacy Imaging LRG sample as a validation of the pipeline under different decontamination schemes.
We considered two decontamination methods depending on the number of maps used to create the weights, \textit{Nonlinear Three Maps} and \textit{Nonlinear Nine Maps}, and focused on the angular range $0.5\deg<\theta<25\deg$ as our optimal configuration.
For the \textit{Nonlinear Nine Maps} case, we found $\fnl^{\rm loc}=-3^{+16}_{-14}$, showing no significant preference for local PNG.
For the \textit{Nonlinear Three Maps} case, we found $\fnl^{\rm loc}=27^{+10}_{-9}$, corresponding to a hint of local PNG, although Rezaie24 showed that this signal is likely due to residual systematics in the decontamination procedure.
For the sample without weights, we found a spurious signal, $\fnl^{\rm loc}=136^{+11}_{-13}$, induced by observational systematics.
These constraints are very similar to the equivalent harmonic space analysis from Rezaie24, further validating our pipeline.

We then analyzed the DESI Imaging data in the extended parameter space including $\Delta$. For \textit{Nonlinear Nine Maps}, the absence of a significant local PNG preference produces a strong $\fnld$--$\Delta$ degeneracy, yielding $\fnld=-0.44^{+1.68}_{-2.17}\times10^{4}$ (68\% C.L.) and the one-sided $95\%$ limit $\Delta>0.83$. 
The resulting one-sided limit on $\Delta$ should therefore not be interpreted as evidence for large $\Delta$.
The unweighted sample, by contrast, gives an apparent constraint, $\Delta=0.22^{+0.07}_{-0.08}$, confirming that observational systematics can generate apparently constrained beyond-local PNG posteriors. 
For \textit{Nonlinear Three Maps}, the local PNG hint maps into a posterior preference for $\fnld=5.12^{+6.13}_{-3.62}\times10^{3}$ and $\Delta=0.91^{+0.25}_{-0.19}$.
This is the first application of the beyond-local PNG model to real galaxy-clustering data with an apparent local-PNG hint, although the strong mitigation dependence prevents interpreting this preference as a signature of massive fields during inflation.

We presented, for the first time, the forecasted sensitivity of Stage-V spectroscopic surveys to massive-field signatures using high-redshift Lyman-break galaxy samples.
We first studied the expected local PNG constraining power for WST, MUST, and Spec-S5.
For the fiducial local PNG $\fnl^{\rm fid}=4$, we found $\sigma(\fnl^{\rm loc})=0.58$ for WST, $\sigma(\fnl^{\rm loc})=0.84$ for MUST, and $\sigma(\fnl^{\rm loc})=1.06$ for SpecS5.
We confirmed that Stage-V surveys can detect local PNG with high significance if the true amplitude is close to current constraints.

To test whether the high local PNG detectability can translate into constraints on massive fields, we first performed a forecast of $\Delta$ around the local PNG limit.
For $\fnldfid=4$ and $\Delta^{\rm fid}=0$, the free-bias constraints are $\sigma(\Delta)=0.17$ for WST, $\sigma(\Delta)=0.34$ for MUST, and $\sigma(\Delta)=0.50$ for SpecS5.
For a stronger local PNG signal, $\fnldfid=18$, the constraints improve to $\sigma(\Delta)\simeq0.06$, $0.09$, and $0.11$, respectively.
We also showed that fixing the galaxy bias parameters improves the constraints on $\Delta$ by about $60\%$.
This indicates that external information on galaxy bias could help break parameter degeneracies and enhance the detectability of $\Delta$.

We found that the constraints on $\Delta$ are related to the local PNG sensitivity.
At low significance, $\text{SNR}_{\rm loc, ref}\lesssim4$, the posterior is affected by the divergent region in the $(\fnld,\Delta)$ plane.
Above this threshold, the contours become more Gaussian and the constraints on $\Delta$ from different surveys follow a common relation, $\sigma(\Delta)\propto \text{SNR}_{\rm loc, ref}^{-1}$.
This provides a useful way to compare the constraining power on $\Delta$ for surveys with different areas, number densities, and redshift ranges.

We then moved beyond the local limit and studied fiducial models with non-zero $\Delta$.
Using WST as a representative Stage-V survey, we showed that the constraints degrade as $\Delta^{\rm fid}$ increases.
This happens because the signal becomes less scale dependent as $\Delta^{\rm fid}$ increases, making the recovery of $\Delta$ progressively harder.
For a sufficiently large PNG amplitude, such as $\fnldfid=16$, WST can recover the fiducial values of both $\fnld$ and $\Delta$ for $\Delta^{\rm fid}=\{0.1,0.4\}$.
In particular, for the free-bias case, we recover $\Delta$ with high significance, but at the expense of degrading the constraints on $\fnld$.
For $\Delta^{\rm fid}=1.0$, we reach the divergent region and cannot recover the fiducial parameters due to prior-volume effects.

We showed that for non-zero $\Delta^{\rm fid}$, the fitted relation between $\sigma(\Delta)$ and $\text{SNR}_{\rm loc, ref}$ remains useful, but it is no longer described by an inverse scaling.
After removing the low significance regime affected by projection effects, we inverted the relation to estimate the local PNG sensitivity required to reach a target uncertainty on $\Delta$.
This translation was calibrated using the fixed-bias configuration and should therefore be interpreted as a best-case requirement.
In the free-bias case, the requirement should become stronger, since the posterior is enlarged by the degeneracy between $\fnld$, $\Delta$, and galaxy bias.

Future work should extend the data analysis by repeating the beyond-local PNG study with a broader set of systematics mitigation methods and with mock catalogs that include signatures of $\Delta$.
This would allow us to quantify how residual systematics project into the massive-field parameter space and to test the robustness of the inferred constraints.

This work showed that future galaxy surveys can not only detect primordial non-Gaussianity, but also identify whether the signal is compatible with the local limit or with signatures of massive fields during inflation, potentially revolutionizing our understanding of the early Universe in the next decade.

\begin{acknowledgments}
The authors thank Mehdi Rezaie for his early contributions to the project, as well as Matteo Braglia and Rogerio Rosenfeld for comments and discussions. 
WR acknowledges support from the S\~ao Paulo Research Foundation (FAPESP), Brasil, with Process Number 2023/07640-7. 
SA and AP have been funded by Agencia Española de Investigación (MCIN/AEI/10.13039/501100011033) under project PID2024-156844NA-C22. SA further received support from the Ramon y Cajal fellowship RYC2022-037311-I. AP also acknowledges support from the \textit{César Nombela} Research Talent Attraction grant from the Community of Madrid (Ref. 2023-T1/TEC-29011).
This research was supported by resources supplied by the Center for Scientific Computing (NCC/GridUNESP) of the São Paulo State University (UNESP).
\end{acknowledgments}

\section*{Data availability}
The DESI Legacy Imaging LRG sample used in this work is publicly available from Ref.~\cite{2024MNRAS.532.1902R} through Zenodo at \url{https://doi.org/10.5281/zenodo.11115914}. 
The remaining data products generated for this analysis are available upon reasonable request.



\bibliography{example} 




\bsp	
\label{lastpage}
\end{document}